\documentclass[twocolumn,prl,showpacs]{revtex4}
\usepackage[colorlinks,bookmarks=false,citecolor=blue,linkcolor=red,urlcolor=blue]{hyperref}
\usepackage{graphicx,epstopdf,color}
\usepackage{amsfonts}
\usepackage{amsmath,amssymb,mathrsfs}
\usepackage{bm}

\newcommand{\bra}[1]{\left\langle#1\right|}
\newcommand{\ket}[1]{\left|#1\right\rangle}

\newcommand{\hc}{\text{H.c.}}
\renewcommand{\i}{\text{i}}

\newcommand{\x}{\text{x}}

\newcommand{\up}{\uparrow}
\newcommand{\dw}{\downarrow}

\newcommand{\vac}{\left|0\right\rangle}

\newlength{\ytlength}
\def\ie{{i.e.},\ }
\def\eg{{e.g.}\ }

\allowdisplaybreaks[1]
\begin{document}
\title{The 1D Ising model and topological order in the Kitaev chain}
\author{Martin Greiter}
\author{Vera Schnells}
\author{Ronny Thomale}
\affiliation{Institute for Theoretical Physics, University of
  W\"urzburg, Am Hubland, 97074 W\"urzburg, Germany}

\pagestyle{plain}

\begin{abstract}
  We elaborate on the topological order in the Kitaev chain, a p-wave
  superconductor with nearest-neighbor pairing amplitude equal to the
  hopping term $\Delta=t$, and chemical potential $\mu=0$.  In
  particular, we write out the explicit eigenstates of the open chain
  in terms of fermion operators, and show that the states as well as
  their energy eigenvalues are formally equivalent to those of an
  Ising chain.  The models are physically different, as the
  topological order in the Kitaev chain corresponds to conventional
  order in the Ising model.
\end{abstract}



\maketitle

{\it Introduction.}---A few years ago, in the lovely town of Trieste,
one of us engaged in a bet with a highly esteemed colleague.  The
issue was whether fermions were physically distinguishable from
hard-core bosons in one dimension (1D), or whether they would only be
different descriptions of the same particles which could be obtained
from each other through gauge transformations.  That they are
distinguishable was settled with the example of two particles on a
ring, where fermions with periodic boundary conditions (PBCs) are
equivalent to hard-core bosons with anti-periodic boundary conditions
(anti-PBCs) and vice versa.  Delivery of the espresso at stake was
promised thereafter.

In this Letter, we provide a much more compelling example of the
difference between fermions and hard-core bosons in 1D.  We will
investigate two simple Hamiltonians, one formulated in terms of
fermions, the other in terms of hard-core bosons realized through
spin-flip operators acting on a Hilbert space with spin
$s=\frac{1}{2}$. Written in a basis of the appropriate operators, the
entire spectrum of eigenstates including their energy eigenvalues is
equivalent for both models.  There is, however, a key difference.  The
states in the fermionic model are topologically
ordered~\cite{wen90ijmpb239,kitaev01pusp131,Wen04,kitaev06ap2,chen-11prb235128,alicea12rpp076501,Bernevig13,bahri-1402.5262},
while the spin model is conventionally ordered in the sense of a
spontaneously broken symmetry.

To be more precise, we investigate the eigenstates of the Kitaev
chain~\cite{kitaev01pusp131,alicea12rpp076501}, a one-dimensional
p-wave superconductor with nearest-neighbor pairing amplitude equal to
the hopping term $\Delta=t$, and chemical potential $\mu=0$, with open
boundary conditions (OBCs).  While those are well known in terms of
the Majorana fermion~\cite{majorana37nc171,wilczek09np614} operators
introduced by Kitaev, we show that they take a very simple yet
somewhat surprising form in terms of the fermion operators which span
the Hilbert space of the model.  We find that both the states and the
Hamiltonian are equivalent to those of an Ising model, with one
crucial difference: The spinless fermion creation and annihilation
operators in the Kitaev model are replaced by bosonic spin flip
operators. 

The ground state of both models
is two-fold degenerate, but the physics of the order displayed could
hardly be more different.  While in the Ising model the $\mathbb{Z}_2$
spin reflection symmetry is spontaneously broken, the degeneracy in
the Kitaev chain stems from the Majorana zero mode (\ie the isolated
Majorana fermions at the ends of the chain) characteristic of the
topological order.

{\it The Kitaev chain.}---Kitaev~\cite{kitaev01pusp131} studied 
a lattice model of a p-wave superconductor in 1D,
\begin{align}
  \label{eq:hpwave}
  H=-\mu\sum_x c_x^\dagger c_x^{\phantom{\dagger}} 
  -\sum_x (t c_x^\dagger c_{x+1}^{\phantom{\dagger}}
  +\Delta e^{\i\phi} c_x^{\phantom{\dagger}}
  c_{x+1}^{\phantom{\dagger}}+\hc ),
  \nonumber\\[-10pt]
\end{align}
where $\mu$ is the chemical potential, $t\ge 0$ the nearest-neighbor
hopping, and 
$\Delta\ge 0$ the p-wave pairing amplitude.  Since the model is
particle hole symmetric, we may restrict our attention to the case
$\mu\le 0$; since the order parameter phase $\phi$ can be absorbed
into the definition of $c_x$ and $c_x^\dagger$, we may set $\phi=0$.
Kitaev showed that this model has two phases: a topologically trivial
strong-coupling phase for $\mu<-2t$, and a topologically non-trivial
weak-coupling phase for $\mu>-2t$.  To understand this, consider first
PBCs and diagonalize \eqref{eq:hpwave} in $k$-space with a standard
Bogoliubov transformation~\cite{deGennes66}.  This yields the
quasiparticle spectrum $\epsilon_k=\sqrt{\xi_k^2+\Delta_k^2}$, where
$\xi_k=-2t\cos k-\mu$, $\Delta_k=2\Delta\sin k$, and we have set the
lattice constant to unity.  The topological order can change only
where the gap closes, which is for $\mu=-t$ at $k=0$.  To illustrate
the two topologically distinct phases, Kitaev turned to a chain with
OPCs, and rewrote the fermion operators in terms of Majorana fermion
operators,
\begin{align}
  \label{eq:gammadef}
  \gamma_{A,x}=-\i c_x^{\phantom{\dagger}} +\i c_x^\dagger,\hspace{10pt}
  \gamma_{B,x}=    c_x^{\phantom{\dagger}} +   c_x^\dagger.
\end{align}
This yields
\begin{align}
  \label{eq:hettore}
  H=&-\frac{\mu}{2}\sum_{x=1}^N (1+\i\gamma_{B,x}\gamma_{A,x}) \nonumber\\
  &-\frac{\i}{2}\sum_{x=1}^{N-1} (\Delta +t) \gamma_{B,x}\gamma_{A,x+1} 
  +(\Delta-t) \gamma_{A,x}\gamma_{B,x+1}). \nonumber\\[-10pt]
\end{align}
The trivial phase is illustrated by the case $t=\Delta=0$, $\mu<0$, in
which Majorana fermions are paired on the same site, and all the sites
are unoccupied.  The topologically non-trivial phase is illustrated by
the case $\mu=0$, $t=\Delta>0$, in which Majorana fermions are paired
on neighboring sites.  This yields an unpaired Majorana fermion at
each end, or a Majorana zero mode formed by combining these two into a
fermion state, which can be occupied or unoccupied.  For OBCs, the
Majorana fermions on the boundaries are a characteristic feature of
the topologically non-trivial phase.  (For PBCs, a characteristic
feature is the fermion parity of the ground state, which is even (\ie
the state consists only of terms with an even numbers of fermions) in
the trivial phase, but odd in the topologically non-trivial phase.
This simple observation seems to have been overlooked in some of the
literature reviewed by Alicea~\cite{alicea12rpp076501}.)

In this Letter, we further investigate the case $\mu=0$, $t=\Delta=1$,
a model we refer to as the Kitaev chain.  The Hamiltonian may be
written
\begin{align}
  \label{eq:Kitaev}
  H_{\text{Kitaev}}&=-\sum_{x=1}^{N-1} 
  (c_{x+1}^\dagger - c_{x+1}^{\phantom{\dagger}}) (c_x^\dagger +
  c_x^{\phantom{\dagger}}) \\
  \label{eq:Kitaevd}
  &=-\i\sum_{x=1}^{N-1} \gamma_{B,x}\gamma_{A,x+1}
  =\sum_{x=1}^{N-1}(2d_x^\dagger d_x^{\phantom{\dagger}}-1)
\end{align}
where 
\begin{align}
  \label{eq:dxdef}
  2d_x^\dagger=\gamma_{B,x}+\i\gamma_{A,x+1}
  =c_{x+1}^{\phantom{\dagger}} -c_{x+1}^\dagger +c_x^{\phantom{\dagger}} + c_x^\dagger .
\end{align}
Closing the OBCs would add another term $(2d_0^\dagger d_0-1)$ to
\eqref{eq:Kitaevd}, where
\begin{align}
  \label{eq:d0def}
  2d_0^\dagger=\gamma_{B,N}+\i\gamma_{A,1}
  =c_{1}^{\phantom{\dagger}} -c_{1}^\dagger +c_N^{\phantom{\dagger}} + c_N^\dagger .
\end{align}
One ground state of \eqref{eq:Kitaevd} is obviously given by the vacuum
defined by the operators $d_x$, $x=0,1,2,\ldots ,N-1$, and the other
is obtained by acting with $d_0^\dagger$ on this vacuum state.  All
the other eigenstates are trivially obtained by creation of various
$d_x^\dagger$ excitations.

{\it Eigenstates in terms of local fermion operators.}---%
It is not obvious, however, how the eigenstates look like in terms of
the original, local fermion operators $c_x$ and $c_x^\dagger$.  A
conceptually straightforward way to obtain them is to choose two seed
states, one with even and one with odd fermion parity, like $\vac$ and
$c_1^\dagger\vac$ (where $c_x\vac=0\ \forall x$), and project them
with
\begin{align}
  \label{eq:P}
  \mathcal{P}\equiv \prod_{x=1}^{N-1}d_x^{\phantom{\dagger}} d_x^\dagger
\end{align}
onto ground states of \eqref{eq:Kitaevd}.  Note that since
\begin{align}
  \label{eq:ddagd}
  2d_x d_x^\dagger=(c_{x+1}^\dagger - c_{x+1}^{\phantom{\dagger}}) 
  (c_x^\dagger + c_x^{\phantom{\dagger}})+1
\end{align}
preserves fermion parity, the projected eigenstates inherit the
fermion parity of the seed states.  For the (unnormalized) ground
states we find (by building up the states site by site and carrying
out the algebra)
\begin{align}
  \label{eq:psievenodd}
  \ket{\psi_0^{\text{even}\atop\text{odd}}}
  =\prod_{x=1}^N(1+c_x^\dagger)\biggl|_{M {\text{even}\atop\text{odd}}}\vac\biggr.,
\end{align}
where $M$ denotes the number of fermion operators in the preceding
product, which we project onto even or odd numbers.  We choose a
convention where products acting on kets are build up from right to
left,
\begin{align}
  \label{eq:ordering}
  \prod_{x=1}^N(1+c_x^\dagger)\equiv
  (1+c_N^\dagger)\cdot\ldots\cdot(1+c_2^\dagger)(1+c_1^\dagger).
\end{align}
For our purposes, it is convenient to introduce an alternative basis
for the two degenerate ground states,
\begin{align}
  \label{eq:psipm}
  \ket{\psi_0^{\pm}}=\prod_{x=1}^N(1\pm c_x^\dagger)
  =\ket{\psi_0^{\text{even}}}\pm\ket{\psi_0^{\text{odd}}}.
\end{align}
We obtain the excited states
\begin{align}
  \label{eq:ddagpsipm}
  d_x^\dagger\ket{\psi_0^{\pm}}
  &=(d_x^\dagger+d_x^{\phantom{\dagger}})\ket{\psi_0^{\pm}}
  =(c_x^{\phantom{\dagger}}+c_x^\dagger)\ket{\psi_0^{\pm}}\nonumber\\
  &=\pm\prod_{y=x+1}^N(1\mp c_y^\dagger)\prod_{y=1}^x(1\pm c_y^\dagger)\vac.
\end{align}
These are just domain walls between the two ground states
$\ket{\psi_0^+}$ and $\ket{\psi_0^-}$.  Trivially, we could have
obtained this result also with
\begin{align}
  \label{eq:ddag-d}
  d_x^\dagger\ket{\psi_0^{\pm}}
  &=(d_x^\dagger-d_x^{\phantom{\dagger}})\ket{\psi_0^{\pm}}
  =(c_{x+1}^{\phantom{\dagger}} -c_{x+1}^\dagger)\ket{\psi_0^{\pm}}.
\end{align}
The terms we sum over in the Hamiltonian \eqref{eq:Kitaev} hence
first create a domain wall between sites $x$ and $x+1$ from one side,
and then annihilate it from the other side.

\emph{Correspondence with the 1D Ising model.}---Since the operators
$d^\dagger_x$ commute for different sites $x$, we can immediately write
down all the eigenstates of \eqref{eq:Kitaev},
\begin{align}
  \label{eq:eigenket}
  \ket{\sigma_1\sigma_2\ldots\sigma_N}\equiv
  \prod_{x=1}^N(1+\sigma_x c_x^\dagger)\vac,
\end{align}
where $\sigma_x=\pm 1$. 
The corresponding energy eigenvalues, defined by
\begin{align}
  \label{eq:Heigenket}
  H\ket{\sigma_1\sigma_2\ldots\sigma_N}
  =E_{\sigma_1\sigma_2\ldots\sigma_N}\ket{\sigma_1\sigma_2\ldots\sigma_N},
\end{align}
are given by
\begin{align}
  \label{eq:Eeigenket}
  E_{\sigma_1\sigma_2\ldots\sigma_N}
  =-\sum_{x=1}^{N-1}\sigma_x\sigma_{x+1}.
\end{align}
The last two equation describe an Ising model in 1D.  We have hence
shown that there is a formal equivalence between the eigenstates 
and energy eigenvalues of the Kitaev model and the Ising model.  

We can make the correspondence more explicit by choosing the Ising
spins in the $x$-direction, while the quantization axis remains the
$z$-axis.  Then the Ising model eigenstates corresponding to
\eqref{eq:eigenket} are given by
\begin{align}
  \label{eq:Isingket}
  \ket{\sigma_1\sigma_2\ldots\sigma_N}\equiv
  \prod_{x=1}^N(1+\sigma_x S_x^+)\ket{\dw}^{\otimes N},
\end{align}
where $\ket{\dw}^{\otimes N}$ denotes a state with all spins $\dw$,
and $S_x^+$ flips a spin at site $x$, $S_x^+\ket{\dw}=\ket{\up}$.
The corresponding Ising Hamiltonian is 
\begin{align}
  \label{eq:Ising}
  H_{\text{Ising}}
  &=-4\sum_{x=1}^{N-1} S_{x+1}^\x S_x^\x\nonumber\\
  &=-\sum_{x=1}^{N-1} (S_{x+1}^+ + S_{x+1}^-) (S_x^+ + S_x^-).
\end{align}
Note that as compared to \eqref{eq:Kitaev}, the sign in the first
factor in \eqref{eq:Ising} is reversed.  This is simply a consequence
of having substituted the fermion operators $c^\dagger$ and $c$ by the
(hard-core) boson operators $S^+$ and $S^-$.  If the site $x+1$ is
occupied in the fermionic model, commuting the factor $(c_x^\dagger +
c_x^{\phantom{\dagger}})$ through it in the state vector we act on
will give us an extra minus sign, which is not present in the bosonic
model.

\emph{Conventional vs.~topological order.}---Irrespective of the
formal equivalence of the two models in the sense elaborated above,
the physical order displayed by them is highly distinct.  The Ising
model displays conventional order, and the ${\mathbb Z}_2$ spin
reflection symmetry $S^\x\to -S^\x$ is spontaneously broken.  There
are no local matrix elements between the two ground states, as one
would have to flip all the spins on the entire chain to transform one
state into the other.  The Kitaev model displays topological order,
and the two-fold ground state degeneracy is due the Majorana
zero-mode, \ie the mode described by the fermion
$d_0^{\phantom{\dagger}},d_0^\dagger$, which consists of the two
Majorana fermions $\gamma_{A,0}$ and $\gamma_{B,N}$ at the end of the
chain.  
In equations,
\begin{align}
  \label{eq:d0psipm}
  (d_0^\dagger-d_0^{\phantom{\dagger}})\ket{\psi_0^{\pm}}
  &=(c_1^{\phantom{\dagger}}-c_1^\dagger)\ket{\psi_0^{\pm}}
  =\pm\ket{\psi_0^{\mp}},
  \nonumber\\
  (d_0^\dagger+d_0^{\phantom{\dagger}})\ket{\psi_0^{\pm}}
  &=(c_N^{\phantom{\dagger}}+c_N^\dagger)\ket{\psi_0^{\pm}}
  =\pm\ket{\psi_0^{\pm}},
\end{align}
and hence
\begin{align}
  \label{eq:d0psiodd}
  d_0^{\phantom{\dagger}}\ket{\psi_0^{\text{odd}}}=0,\hspace{6pt}
  d_0^{\dagger}\ket{\psi_0^{\text{odd}}}=\ket{\psi_0^{\text{even}}}.
\end{align}
The only physical difference between $\ket{\psi_0^{\text{odd}}}$ and
$\ket{\psi_0^{\text{even}}}$ is the occupation of the Majorana-zero
mode, which can easily be altered by creation and annihilation of 
fermions at the boundaries.  These two ground states differ in their
fermion parity, which is only a global, but not a local property.

Interestingly, if we diagonalize both models numerically, and set up
Hilbert space conventions in which at each site $x$ for the Kitaev
model empty (\ie $\vac$) and occupied (\ie $c_x^\dagger\vac$), and for
the Ising model $\dw$-spin (\ie $\ket{\dw}$) and $\up$-spin (\ie
$S_x^+\ket{\dw}$), by 0 and 1, the eigenstates of \eqref{eq:Kitaev}
and \eqref{eq:Ising} would be identical.

This is not to say that the correlations of both models are identical,
or even related.  A correlation function is, like an order parameter,
an expectation value of an operator (or product of
operators) in a ground state.  While we can easily measure the Ising
spin $2S_x^\x=S_x^+ + S_x^-$ on any site $x$ in an eigenstate of
\eqref{eq:Ising},
\begin{align}
  \label{eq:Isingexpval}
  \bra{\sigma_1\sigma_2\ldots\sigma_N} S_x^+ + S_x^-
  \ket{\sigma_1\sigma_2\ldots\sigma_N}=\sigma_x,
\end{align}
there is no corresponding, local operator to measure $\sigma_x$ in an
eigenstate of the Kitaev model \eqref{eq:Kitaev}.  In particular,
\begin{align}
  \label{eq:Kitaevexval}
  \bra{\sigma_1\sigma_2\ldots\sigma_N} c_x^\dagger+c_x^{\phantom{\dagger}}
  \ket{\sigma_1\sigma_2\ldots\sigma_N}=0\quad \forall\; x<N.
\end{align}

It is worth pointing out, however, that the entanglement
spectrum~\cite{peschel03jpa,li-08prl010504}, is identical for the
ground states of both models.  The comparison illustrates that not
only the nature of the cut itself, but also the \mbox{(non-)locality}
of the basis (\ie fermions vs.\ bosonic spin flips operators) in which
the reduced density matrix is formulated, must be taken into account
when interpreting the entanglement spectrum.

\emph{Reconciliation with the BCS pairing wave function.}---%
We now wish to reconcile our ground state wave function
\eqref{eq:psievenodd} for the Kitaev's p-wave superconductor
\eqref{eq:Kitaev} with the conventional form of a BCS wave function in
position space.  To begin with, let us take another look at our wave
function.  As we close the OBCs by adding a term $(2d_0^\dagger
d_0-1)$ to \eqref{eq:Kitaevd}, the ground state becomes non-degenerate
and is given by $\ket{\psi_0^{\text{odd}}}$ (see \eqref{eq:d0psiodd}).
Note that if we reinstate the phase $\phi$ in \eqref{eq:hpwave} which we
absorbed into the definition of $c_x^\dagger$ and
$c_x^{\phantom{\dagger}}$, we may write the ground state
as
\begin{align}
  \label{eq:psiodd}
  \ket{\psi_0^{\text{odd}}(\phi)}
  &=\prod_{x=1}^N
  (1+e^{-\frac{\i}{2}\phi} c_x^\dagger)\Bigl|_{M\, {\text{odd}}}\vac\Bigr.,\\
  \label{eq:psioddpm}
  &=\pm\prod_{x=1}^N
  (1\pm e^{-\frac{\i}{2}\phi} c_x^\dagger)\Bigl|_{M\, {\text{odd}}}\vac\Bigr.,
\end{align}
At first sight, this may look like a BCS wave function for the
condensation of single fermions rather than Cooper pairs.  This is of
course misguided, as there is no order parameter associated with the
phase between the two terms in \eqref{eq:psiodd}.  At the same time,
it doesn't look much like the wave function of a superconductor, and does
not allow us to read off the Cooper pair wave function directly.  (On
a side note, \eqref{eq:psiodd} shows that a rotation of the
superconducting order parameter phase in \eqref{eq:hpwave} maps onto a
rotation of the Ising spin axis in the $xy$-plane in
\eqref{eq:Ising}.)

To obtain the Cooper pair wave function, we go back to the Kitaev
Hamiltonian \eqref{eq:Kitaev}, and solve it via a standard Bogoliubov
transformation in momentum space.  This yields
\begin{align}
  \label{eq:psiBCS}
  \ket{\psi_0}=\prod_{0<k<\pi}
  (u_k + v_k c_k^\dagger c_{-k}^\dagger)\cdot c_{k=0}^\dagger\vac ,
\end{align}
where the product extends over all discrete $k=\frac{2\pi}{N}n$
(with $n$ integer) in the specified interval, $u_k=\sin\frac{k}{2}$,
and $v_k=-\i\cos\frac{k}{2}$.  Leaving aside the overall normalization,
we may rewrite \eqref{eq:psiBCS} as (see \eg \cite{greiter05ap217},
App.~A)
\begin{align}
  \label{eq:psiBCS1}
  \ket{\psi_0}=\exp(b^\dagger)\cdot c_{k=0}^\dagger\vac,
\end{align}
where
\begin{align}
  \label{eq:bk}
  b^\dagger=\sum_{0<k<\pi} \frac{v_k}{u_k} c_k^\dagger c_{-k}^\dagger.
\end{align}
creates a Cooper pair.  Transforming this into position space, we
obtain
\begin{align}
  \label{eq:bxy}
  b^\dagger=\sum_{x>x'} \varphi_{x-x'}^{\phantom{\dagger}} 
  c_{x\phantom{'}}^\dagger c_{x'}^\dagger
\end{align}
with
\begin{align}
  \label{eq:varphi}
  \varphi_{x-x'}^{\phantom{\dagger}} 
  =\frac{1}{N}\sum_{k\ne 0} \frac{v_k}{u_k} e^{\i k(x-x')}
  =1-\frac{2(x-x')}{N},
\end{align}
where we have evaluated the sum for $0<x-x'<N$ using (see \eg
\cite{Greiter11}, App.~B)
\begin{align}
  \label{eq:app-hsfouriersum1}
  \sum_{\alpha=1}^{N-1}\frac{\eta_\alpha^n}{\eta_\alpha -1}
  =\frac{N+1}{2}-n,\hspace{8pt} 
  \eta_\alpha\equiv e^{\i\frac{2\pi}{N}\alpha},
\end{align}
which holds for $1\le n \le N$. 

The analysis presented so far implies that \eqref{eq:psiodd} (with
$\phi=0$) and \eqref{eq:psiBCS1} with \eqref{eq:bxy} and
\eqref{eq:varphi} are equivalent.  As this is not obvious to the eye,
we now show it explicitly by comparing  terms with the same number of
fermions $M$ in
\begin{align}
  \nonumber \exp(b^\dagger)\cdot c_{k=0}^\dagger\vac
  \hspace{8pt} \text{and}\hspace{8pt}
  \prod_{x=1}^N (1+c_x^\dagger)\Bigl|_{M\, {\text{odd}}}\vac\Bigr..
\end{align}
Since
\begin{align}
  \label{eq:psiM}
  \prod_{x=1}^N (1+c_x^\dagger)\Bigl|_{M}\vac\Bigr.
  =\hspace{-8pt} \sum_{y_M>\ldots >y_2>y_1\rule{0pt}{8pt}}\hspace{-8pt} 
  c_{y_M}^\dagger\ldots c_{y_2}^\dagger c_{y_1}^\dagger \vac,
\end{align}
it is sufficient to show that 
\begin{align}
  \label{eq:allm}
  \bra{0} c_{y_1}c_{y_2}\ldots c_{y_M} 
  \left(b^\dagger\right)^m\sum_{x_1} c_{x_1}^\dagger\vac = m!,
\end{align}
where 
$m=(M-1)/{2}$ is the number of Cooper pairs, 
and $y_1<y_2<\ldots <y_M$.  As \eqref{eq:allm} holds trivially for
$M=1$, all we have to show to complete the proof inductively is that
\begin{align}
  \label{eq:tobeshown}
  &\bra{0} c_{y_1}\ldots c_{y_M}\, 
  b^\dagger \hspace{-8pt}\sum_{x_{M-2}>\ldots >x_1} \hspace{-8pt}
  c_{x_{M-2}}^\dagger\ldots c_{x_1}^\dagger\vac\nonumber\\ 
  &=\bra{0} c_{y_1}\ldots c_{y_M}\, 
  \hspace{-8pt}\sum_{x_{M-2}>\ldots >x_1} \hspace{-8pt}
  c_{x_{M-2}}^\dagger\ldots c_{x_1}^\dagger\, b^\dagger \vac = m
\end{align}
holds for $M\ge 3$, $y_j<y_{j+1}$, and $b^\dagger$ given by
\eqref{eq:bxy} and \eqref{eq:varphi}.  In evaluating
\eqref{eq:tobeshown}, we first consider the contribution of the second
term in \eqref{eq:varphi}.  When we order all the site indices
$x',x,x_1,\ldots x_{M-2}$ in ascending order, let $x'$ be number $i'$
and $x$ number $i$ in the list.  For a given $y_j$ to contribute
$-\frac{2}{N}y_j$ in \eqref{eq:tobeshown}, either $x$ or $x'$ has to
be equal to $y_j$.  For $x=y_j$, $x'$ has to be equal to a smaller
$y$, and hence all values $i'\in [1,j-1]$ will contribute with sign
$(-1)^{j+i'+1}$.  Similarly, for $x'=y_j$, all values $i\in [j+1,M]$
will contribute with sign $(-1)^{i+j+1}$.  The overall contribution
$\propto y_j$ is hence
\begin{align}
  \label{eq:zero}
  -\frac{2}{N}y_j \Biggl\{
  \sum_{i'=1}^{j-1} (-1)^{j+i'+1} - \sum_{i=j+1}^{M} (-1)^{i+j+1}
  \Biggr\}=0.
\end{align}
This leaves us with the first term in \eqref{eq:varphi}, which by
a similar argument yields
\begin{align}
  \label{eq:firstterm}
  \sum_{i>i'}^{M} (-1)^{i+i'+1}=m.
\end{align}
This completes the proof.

\emph{Mapping by Jordan--Wigner-transformation.}---It has been noted
previously that the models \eqref{eq:Kitaev} and \eqref{eq:Ising} can
be transformed into each other via a
Jordan--Wigner-transformation~\cite{kitaev-0904.2771,bardyn-12prl253606,zvyagin13prl217207}.
These studies, however, rely on the assumption 
that the Ising model has topological order as well, and that a
Majorana fermion zero-mode can hence also be observed in bosonic
models. One result of our study is that this is not possible.

\emph{Conclusion.}---We have demonstrated that a fermion model with
topological order, the 1D $p$-wave superconductor studied by Kitaev,
can (as far as eigenstates and their energies are concerned) be mapped
into a boson model with conventional order, the 1D Ising model. 
This suggests that other models with topological order, such as
Kitaev's toric code or honeycomb model in
2D~\cite{kitaev03ap2,kitaev06ap2}, might have simpler, bosonic cousins
with conventional order.  Inversely, reformulating certain bosonic
models with conventional order due to a broken discrete symmetry, in
terms of fermion operators, may provide a route to novel models with
topological order.

{\it Acknowledgments.}---We wish to thank T.~Neupert, S.~Rachel, and
D.~Schuricht for bringing
references~\cite{kitaev-0904.2771,bardyn-12prl253606,zvyagin13prl217207}
to our attention. This work was supported by the ERC starters grant
TOPOLECTRICS under ERC-StG-Thomale-336012.




\vfill\eject
\end{document}